\documentstyle[prb,twocolumn,aps]{revtex}
 \sloppypar
\begin{document}
\draft

\twocolumn
[\hsize\textwidth\columnwidth\hsize\csname
@twocolumnfalse\endcsname

\title{Temperature Evolution of the Quantum Gap in ${\bf CsNiCl_{3}}$}

\author{M. Kenzelmann$^{1}$, R.~A. Cowley$^{1}$, W.~J.~L. Buyers$^{2,3}$, D.~F. McMorrow$^{4}$}
\address{$^{1}$
Oxford Physics, Clarendon Laboratory, Oxford OX1 3PU, United Kingdom} %
\address{$^{2}$
Neutron Program for Materials Research, National Research Council
of Canada, Chalk River, Ontario, Canada KOJ 1J0} %
\address{$^{3}$
Canadian Institute for Advanced Research} %
\address{$^{4}$
Condensed Matter Physics and Chemistry Department, $Ris\o$
National Laboratory, DK-4000, Roskilde, Denmark}%

\date{\today}
\maketitle
\begin{abstract}
Neutron scattering measurements on the one-dimensional gapped S=1
antiferromagnet, ${\rm CsNiCl_{3}}$, have shown that the
excitation corresponding to the Haldane mass gap $\Delta$ at low
temperatures persists as a resonant feature to high temperatures.
We find that the strong upward renormalisation of the gap
excitation, by a factor of three between $5$ and $70\;\mathrm{K}$,
is more than enough to overcome its decreasing lifetime. We find
that the gap lifetime is substantially shorter than that predicted
by the scaling theory of Damle and Sachdev in its low temperature
range of validity. The upward gap renormalisation agrees with the
non-linear sigma model at low temperatures and even up to $T$ of
order $2\Delta$ provided an upper mass cutoff is included.
\end{abstract}

 ]

\newpage

The excitations of one-dimensional (1D) Heisenberg
antiferromagnets have attracted much experimental and theoretical
attention ever since Haldane \cite{F_Haldane} predicted that the
excitations of integer spin and half-integer spin chains are
different. It is now well established that for half-integer chains
the excitations have a continuous spectrum extending to zero
energy while for integer spin chains there is an energy gap at low
temperatures. How the excitations evolve into quasielastic
paramagnetic scattering in the high temperature limit is unknown.
Recently, the temperature dependence of the gap excitation for an
integer spin chain has been discussed by Damle and Sachdev (DS)
\cite{Damle_Sachdev} using the $O(3)$ sigma model of the low
temperature excitations. They derive an expression for the
temperature dependence of the inelastic neutron scattering
line-shape that should apply to all 1D systems that have a gap
$\Delta$ in the excitation spectrum at low temperatures, $k_{\rm
B} T < \Delta$. They also derive a scaling form for the scattering
in energy, momentum and temperature. We have measured the
temperature dependence of the neutron scattering and compare the
results with the theory \cite{Damle_Sachdev}. Good agreement is
obtained with the predicted line-shape, but the theory fails to
describe the observed lifetimes of the excitations. Our results
further show that the excitation has a resonant form for high
temperatures $T$ of order of the bandwidth of the excitation $2.7
J$ or $6 \Delta$.\par

The experiments were performed using the quasi-1D antiferromagnet,
${\rm CsNiCl_{3}}$. Previous neutron scattering experiments have
shown that the spin excitations of its disordered phase above
$4.85\;\mathrm{K}$ have an energy gap as predicted by Haldane for
$S$=$1$ chains \cite{William_Buyers_1,Rose_Morra}, while the
excitation is a spin triplet \cite{Michael_Steiner,Z_Tun_2,Z_Tun},
and the dispersion of the excitations has $2\pi$ periodicity
showing that the symmetry of the spin chain is not broken
\cite{Z_Tun_2,Z_Tun}. All of these properties are consistent with
the properties expected for a quantum disordered phase close to a
quantum critical point. The bare exchange interaction $J$ along
the hexagonal c-axis of the chains is $2.28\;\mathrm{meV}$
\cite{Z_Tun,Katori}, the inter-chain interaction $J'$ is
$0.044\;\mathrm{meV}$ and the Hamiltonian can be written as
\begin{equation}
H=J \sum_{i}^{\rm chain} \vec{S}_{\rm i} \cdot \vec{S}_{\rm i+1}+
J' \sum_{<i,j>}^{\rm plane} \vec{S}_{\rm i} \cdot \vec{S}_{\rm j}
- D \sum_{i} (\vec{S_{\rm i}^{z}})^2\, . \label{Hamiltonian}
\end{equation}
The single-ion anisotropy $D$ of $4\;\mathrm{\mu eV}$ is small so
that ${\rm CsNiCl_{3}}$ is a good example of an $S$=$1$ isotropic
antiferromagnetic (AF) chain system \cite{Rose_Morra,M_Enderle_1}.
The upper limit of the excitation band, $6\;\mathrm{meV}$, is
$30\%$ larger than $2J$ because of quantum renormalisation
\cite{Z_Tun}. Below $4.85\;\mathrm{K}$ the inter-chain
interactions give rise to three-dimensional (3D) long range AF
order \cite{Clark}. In this 3D phase longitudinal and transverse
excitations have been observed and arise from the breaking of the
triplet symmetry of the high temperature excitations
\cite{M_Enderle_1}.\par

A single crystal of ${\rm CsNiCl_{3}}$ $20\;\mathrm{mm} \times
5\;\mathrm{mm} \times 5\;\mathrm{mm}$ was mounted in a cryostat
with its (hhl) crystallographic plane in the horizontal scattering
plane of the neutron scattering instrument, and  the temperature
was controlled with an accuracy of $0.1 \;\mathrm{K}$ between
$1.5$ and $70\;\mathrm{K}$. The experiments were performed with
the RITA triple axis crystal spectrometer \cite{Mason} at the DR3
reactor of the ${\rm Ris\o}$ National Laboratory. The energy of
the neutrons thermalised in a cold source was selected by a
vertically focusing pyrolytic graphite monochromator. A rotating
velocity selector before the monochromator suppressed unwanted
neutrons that would be Bragg reflected by its higher order planes.
Supermirror guides with $n=3$ ($\theta_{\rm c} =1.2^{\rm o}$ at
$4\;\mathrm{\AA}$) were located before and after the
monochromator, and the beam was $20\;\mathrm{mm}$ wide. The
scattered neutron beam was filtered through cooled beryllium,
passed through a $50\;\mathrm{mm}$ wide $1^{\rm o}$ Soller
collimator, and analysed by reflection from the central two blades
of a 7-component flat pyrolytic graphite analyser aligned so that
each blade reflected the same energy of neutrons,
$5\;\mathrm{meV}$. The Soller geometry meant that the beam
reflected by the analysers was about 20 mm wide and located in the
central 30 strips of a position sensitive detector. Because the
detector was 120 pixels wide ($\sim120\;\mathrm{mm}$) the side
strips were used to estimate the temperature independent
background, which was subtracted from the data. Turning of either
analyser or monochromator from their reflecting position confirmed
that the side pixels gave a good representation of the background
arriving at the centre of the detector with the signal. With this
arrangement the low-energy resolution was typically
$0.26\;\mathrm{meV}$ (FWHM) in energy. The longitudinal and
transverse wave-vector width of the $(002)$-peak was $0.01$ and
$0.016$ (FWHM in reciprocal lattice units) and the calculated
vertical resolution was $0.22\;\mathrm{\AA^{-1}}$.\par

Measurements were made for wave-vector transfers along the $[\eta
\eta 1]$ direction, which corresponds to a $\pi$ phase difference
between the Ni spins along the chains. In the absence of the
interactions between the chains, the energy of the spin
excitations would be independent of the wave vector perpendicular
to the chain direction or c-axis. There is however a significant
dispersion that is proportional to the Fourier transform of the
inter-chain exchange interactions. The minimum of the dispersion
relation along the $[\eta \eta 1]$ direction occurs when
$\eta=1/3$, which is the ordering wave vector of the low
temperature AF structure. For most of our measurements we set
$\eta=0.81$, where the Fourier transform of the inter-chain
interactions is zero, so that to first order our results are
characteristic of independent $S$=$1$ chains. Scans were performed
by varying the energy transfer between $0$ and
$13.5\;\mathrm{meV}$, and were repeated for 16 temperatures
between $1.5$ and $70\;\mathrm{K}$.\par

The results for two temperatures are shown in Fig.~\ref{twoscans}.
It is clear that both the energy and the width of the excitations
increase rapidly with increasing temperature. Surprisingly we find
that the excitations are reasonably defined up to
$70\;\mathrm{K}$.
\par
\begin{figure}
  \caption{Neutron scattering intensity at $9\;\mathrm{K}$ and
$50\;\mathrm{K}$ at (0.81 0.81 1) wave-vector transfer as a
function of energy transfer. The energy resolution $\Delta
=0.26\;\mathrm{meV}$ is shown as a horizontal bar. The solid line
is a fit of a Gaussian to the quasi-elastic peak and a double
Lorentzian for the magnetic excitation as explained in the text.}
  \label{twoscans}
\end{figure}

The experimental results were analyzed by fitting the results at
low temperature, $1.5 \;\mathrm{K}$, to a Lorentzian form, which
gave a better fit than the Gaussian form. At this temperature the
magnetic structure is an ordered antiferromagnet and the
excitation is expected to be a long-lived spin wave. We therefore
assume that the width of $0.26\;\mathrm{meV}$ at $1.5\;\mathrm{K}$
is a good measure of the experimental resolution function. As the
temperature is raised in the 1D phase the response is found to
have a Lorentzian spectral form (Fig.~\ref{twoscans}). This is in
agreement with the DS prediction \cite{Damle_Sachdev} based on a
model of collisions between the injected $Q=\pi$ spin excitation
and a gas of thermally activated excitations near the bottom of
the gapped band. The gas particles were given classical
dispersion. Our data was described by a constant background, a
quasi-elastic Gaussian for the incoherent scattering, and, for the
excitation, a double Lorentzian form:
\begin{eqnarray}
        S(E)= A \cdot \left( n(E)+1 \right) \cdot \hspace{1cm} \nonumber \\
        \left(\frac{\Gamma}{(E-\epsilon(0))^2 +\Gamma^2} -
        \frac{\Gamma}{(E+\epsilon(0))^2+\Gamma^2}\right)\, ,
\end{eqnarray}
where $E$ is the energy transfer, $n(E)$ is the Bose factor, the
energy width is given by $\Gamma$ and the gap energy by
$\epsilon(0)$ and $A$ is a constant. The parameters $\Gamma$,
$\epsilon(0)$ and $A$ were fitted at each temperature and, as
shown in Fig.~\ref{twoscans}, the model gave a good description of
the data. The results for the widths are shown in
Fig.~\ref{widths} and the resonant energies $\epsilon(0)$ are
shown in Fig.~\ref{gap}. Both show a rapid initial increase with
increasing temperature. By a temperature of about $2J/k_{\rm B}$
the energy has increased by more than a factor of 3 and the
excitation is still well defined. We have found that a
non-resonant overdamped Lorentzian cannot account for the data at
any temperature. This confirms that the excitations remain as a
resonance to high temperatures.\par

The temperature dependence of the excitations in ${\rm
CsNiCl_{3}}$ has also been studied by Zaliznyak et al
\cite{I_A_Zaliznyak}. Their measurements, however, were at the 3D
minimum of the excitation energy at a wave-vector transfer of
(0.33,0.33,1) and for temperatures below $20 \;\mathrm{K}$. They
observed an increase in the line-width and energy with increasing
temperature but their results are strongly influenced by 3D
effects and so cannot be directly compared with the theory of
quantum disordered chains.\par

The initial motivation for these experiments was the theory of the
neutron scattering cross-section in 1D gapped antiferromagnets by
DS \cite{Damle_Sachdev}. They suggested that for $k_{\rm B}T <
\Delta$ the cross-section is given by a scaling from:
\begin{equation}
S(Q,E) = \frac{Ac}{\Gamma \Delta} \Phi \left( \frac{E-\epsilon(q)}
{\Gamma}\right) \label{structure_factor}
\end{equation}
where $A$ is a constant. They described the dispersion of the
excitations by a classical approximation to the non-linear sigma
model (${\rm NL\sigma M}$) namely:
\begin{equation}
\epsilon(q) = \Delta + \frac{c^2q^2}{2\Delta}\, .
\label{dispersion}
\end{equation}
The difference between the wave-vector transfer, $Q$, and the AF
wave vector is $Q-\pi = q$.  The energy broadening is proportional
to the density of excited excitations times their root mean square
velocity, which DS obtain analytically as
\begin{equation}
L_{\rm t}^{-1}=\Gamma = \frac{3 k_{\rm B}T}{\sqrt{\pi}}
\exp\left(-\frac{\Delta}{k_{\rm B}T}\right)\, ,
\label{theorie_width}
\end{equation}
where $L_{\rm t}$ is the excitation lifetime. The scaling function
$\Phi(z)$ was calculated numerically and is found to be given
closely by a Lorentzian form:
\begin{equation}
\Phi(z) = \frac{\pi \alpha}{2(\alpha^2+z^2)} \label{function}
\end{equation}
with the constant $\alpha=0.71 \sim 1/\sqrt{2}$. Thus the theory
predicts that the observed half-width should be $\alpha \cdot
\Gamma$, with $\Gamma$ from Eq.~\ref{theorie_width}.\par

In the case of ${\rm CsNiCl_{3}}$, the energy gap at absolute zero
cannot be reliably obtained because it enters the 3D phase
transition below $4.85\;\mathrm{K}$. The best estimate is obtained
from the exchange constant \cite{Katori}, in agreement with an
earlier estimate \cite{Z_Tun}, using the relation
$\Delta_{0}=0.41J$ \cite{Minoru_Takahashi_1} to give
$\Delta_{0}=0.93 \;\mathrm{meV}\simeq 11\;\mathrm{K}$.\par

\begin{figure}
  \caption{The half-width, $\Gamma$, of the excitation.
The points are the experimental half-widths and the solid line is
the ${\rm NL\sigma M}$ prediction by DS. The inset shows the
widths for $T < 15\;\mathrm{K}$. The dashed line is our
modification of the DS theory to include the relativistic
dispersion.}
  \label{widths}
\end{figure}

In Fig.~\ref{widths} we compare the half-width of the excitations
with the prediction of $\alpha \cdot \Gamma$ from
Eq.~\ref{theorie_width} and \ref{function} with
$\Delta=\Delta_{0}$, the gap at $T=0\;\mathrm{K}$. For low
temperatures, $k_{\rm B}T < \Delta$, where the theory is
applicable, the theoretical widths are much less than the observed
widths. Around $10\;\mathrm{K}$, the experimental widths are about
$40\%$ times the predicted width. This discrepancy may be due
indirectly to 3D interactions which are still appreciable in this
temperature range. The energy of the spin excitation measured at
the 1D point, (0.81 0.81 1), up to $15\;\mathrm{K}$ are noticeably
larger than at the 3D ordering wave vector
\cite{Michael_Steiner,I_A_Zaliznyak}, while it is lower than at
other wave-vectors \cite{Rose_Morra,Michael_Steiner}, but the
average effect may still not fully cancel at the 1D point we have
selected to study. However, good agreement between theoretical and
experimental width is achieved, we have found (Fig.~\ref{widths}),
if the classical dispersion in DS's model is replaced by the
relativistic dispersion of the non-linear sigma model ${\rm
NL\sigma M}$ ($\epsilon(q) = (\Delta^2 + c^2 q^2)^{1/2}$). The
calculation of $\Delta$ must be done numerically. At temperatures
above about $20\;\mathrm{K}$, the rate of increase in the observed
width decreases, an effect which is not reproduced by the theory.
In this temperature region the ${\rm NL\sigma M}$ is no longer
expected to apply. We find that including the temperature
dependence of the excitation energy does not improve the agreement
of the theory.\par

\begin{figure}
  \caption{The measured excitation energy is compared with the
theoretical predictions. The solid line is the prediction of ${\rm
Jolic\oe ur}$ from the ${\rm NL\sigma M}$ and the dashed line is
the model of K\"{o}hler and Schilling. The inset shows the gap for
$T<15\;\mathrm{K}$ compared with the prediction of the ${\rm
NL\sigma M}$.} \label{gap}
\end{figure}

The energy of the excitation shown in Fig.~\ref{gap} increases by
a factor of 3.8 from $1.22 \;\mathrm{meV}$ at $5\;\mathrm{K}$ to
$4.6\;\mathrm{meV}$ at $70 \;\mathrm{K}$. This substantial
increase in the excitation energy of a gapped quantum chain has
not been previously reported and contrasts with the decrease seen
in most two- and three-dimensional antiferromagnets. This fast
upward renormalisation is the reason that the excitation remains
resonant to high temperatures, much higher than observed in
earlier surveys of the temperature dependence of the excitation
energy \cite{William_Buyers_4}. The measured energy at
$9\;\mathrm{K}$ agrees well with the original estimate of the
energy \cite{William_Buyers_1}.\par

The upward renormalisation of the excitation energy is compared in
Fig.~\ref{gap} with the calculation based on the self-consistent
${\rm NL\sigma M}$ theory of ${\rm Jolic\oe ur}$ and Golinelli
\cite{Jolicoeur_Golinelli} using $\Delta_{0}=0.93\;\mathrm{meV}$
and a relativistic dispersion. The theory describes the measured
energy quite well between 8 and $15\;\mathrm{K}$, where it
predicts a slightly lower gap energy than observed. The difference
of a few percent lies within the error of $J$, with which the
predicted gap energy scales for all $T$. The upward gap
renormalisation is not included in the theory by DS, although
their theory should be valid for $k_{\rm B}T < \Delta =
11\;\mathrm{K}$.\par

Below $6\;\mathrm{K}$, the observed energy flattens off with
decreasing temperature. This is possibly because, as the ${\rm
N\acute{e}el}$ temperature is approached, the interactions between
the chains reduce the effective excitation energy so that the gap
must become larger than for a pure 1D system, so as to conserve
the total spin. The greater thermal excitation increases the gap
energy above that of the purely 1D ${\rm NL\sigma M}$.\par

The decreasing temperature dependence of the measured excitation
energies above $50\;\mathrm{K}$ contrasts with the steadily
increasing gap energy of the ${\rm NL\sigma M}$. This partly
reflects the assumption of an infinitely large distribution of
possible energies in the continuum ${\rm NL\sigma M}$. If we
constrain the upper momentum to $2\pi$ because of the lattice, the
energies tend to level off at higher temperatures. It is
recognized, however, that the mapping of the 1D Heisenberg
Hamiltonian to the ${\rm NL\sigma M}$ is not applicable at such
high temperatures.\par

K\"{o}hler and Schilling \cite{Koehler_Schilling} have described
the temperature dependence of the excitations in terms of a
restricted set of zero-spin defect states and treated them with a
Hartree-Fock approximation. Their result describes the energy very
well for $T < 20\;\mathrm{K}$ (Fig.~\ref{gap}), but only if
$J=2.8\;\mathrm{meV}$ is used as fit parameter. At higher
temperatures, the theory predicts a levelling off of the energy as
observed, but it underestimates the energy in the high temperature
limit. The upturn of the energy at low temperatures may be an
artefact of the theory.\par

The excitation remains resonant up to at least $70\;\mathrm{K}
\sim 2.7 J$. This is more than twice the temperature of the
maximum in the magnetic susceptibility and in the specific heat
\cite{N_Achiwa,Moses}, which has previously been taken to indicate
the breaking of the 1D correlations. Our results are in sharp
contrast with a quantum Monte Carlo calculation where the peak in
the spectral function persists only to $0.5 J \sim 13\;\mathrm{K}$
\cite{Deisz_Cox}.\par

In summary, we find\\1. that the excitations for the AF
wave-vector persist as a resonant feature at temperatures greater
than the spin band energy;\\2. that a scaling form of the theory,
in its expected range of validity, overestimates the lifetime of
the excitations by $40\%$;\\3. that the upward renormalisation of
the energy of the excitations predicted by the ${\rm NL\sigma M}$
theory is accurate only in a narrow range and fails at high
temperatures.\par

A theory is needed that includes the persistence of the resonance,
the strong temperature dependence of the excitation energy, and
the correct line-width.\par

\vspace{0.2cm}

We are grateful to I. Affleck, T. ${\rm Jolic\oe ur}$ and M.
Enderle for discussions. Financial support for the experiments was
provided by the EPSRC, by the EU through its Large Installations
Program and by the British Council-NRC Program. One of us (M.~K.)
is supported by a TMR-fellowship from the Swiss National Science
Foundation under contract no. 83EU-053223.

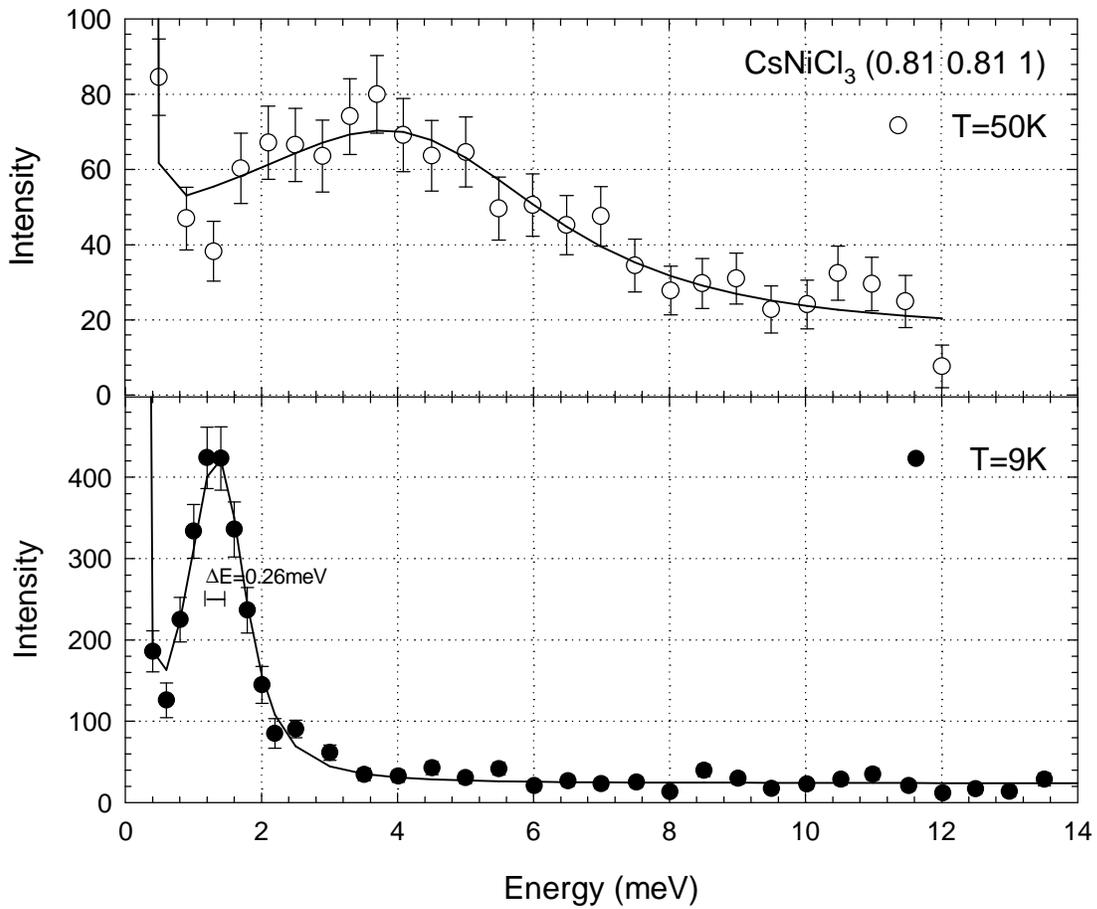
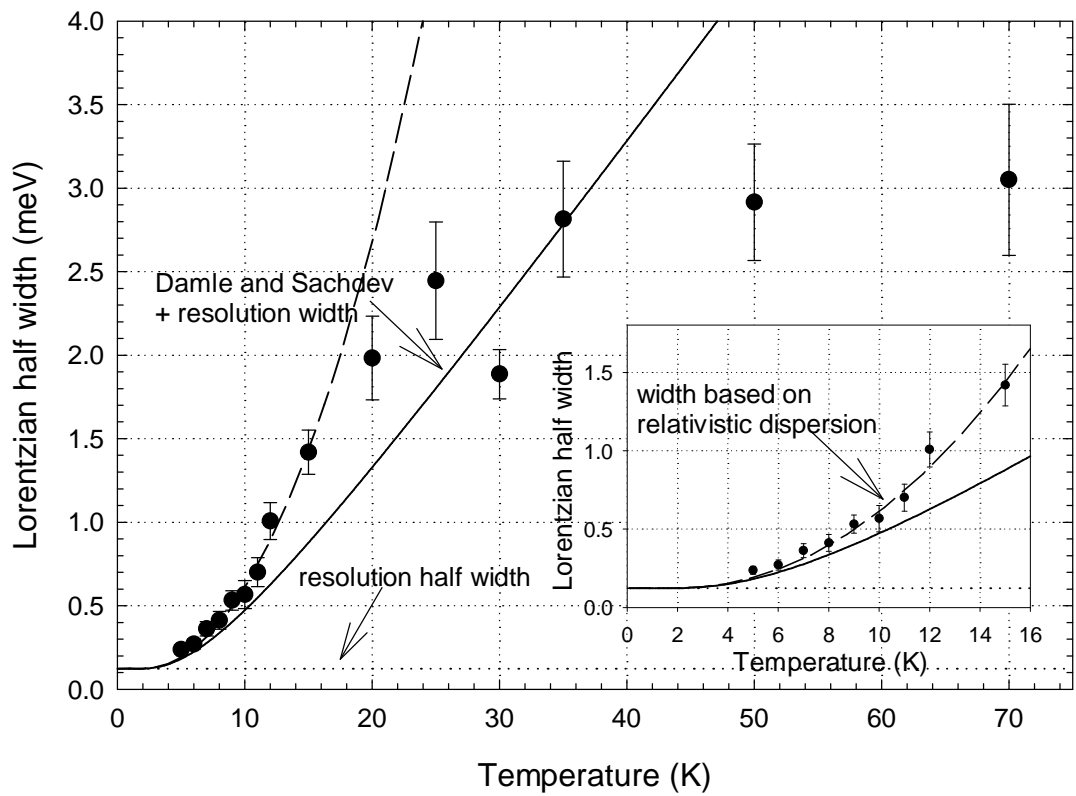
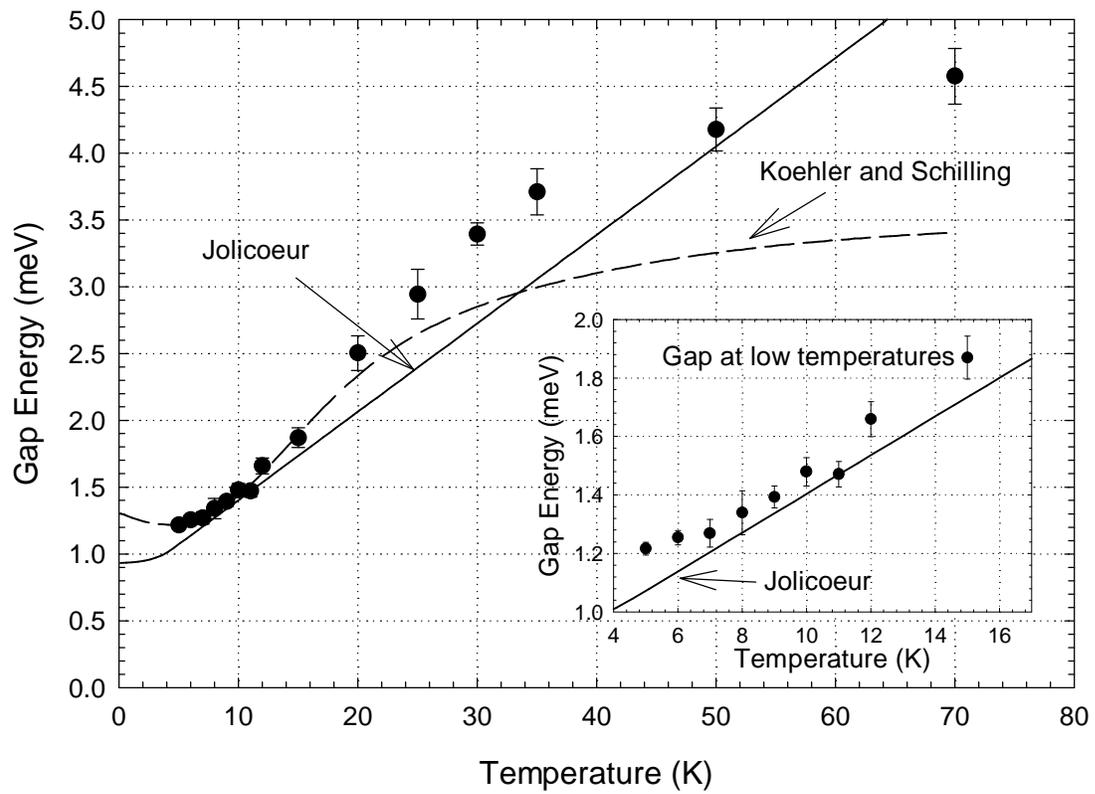
\end{document}